\documentstyle[11pt,aaspp4,epsfig]{article}
\slugcomment{The Astrophysical Journal Letters: 534, L31, 2000 May 1}

\begin{document}

\title{On the Disappearance of Kilohertz Quasi-Periodic Oscillations 
at a High Mass Accretion Rate in Low-Mass X-ray Binaries}
\author{Wei Cui} 
\affil{Center for Space Research, Massachusetts Institute of Technology, 
Room 37-571, Cambridge, MA 02139; cui@space.mit.edu}

\begin{abstract}
For all sources in which the phenomenon of kilo-Hertz quasi-periodic 
oscillation (kHz QPO) is observed, the QPOs disappear abruptly when
the inferred mass accretion rate exceeds a certain threshold. Although
the threshold cannot at present be accurately determined (or even
quantified) observationally, it is clearly higher for bright Z sources 
than for faint atoll sources. Here we propose that the observational 
manifestation of kHz QPOs {\em requires} direct interaction between 
the neutron star magnetosphere and the Keplerian accretion disk and 
that the cessation of kHz QPOs at high accretion rate is due to 
the lack of such an interact when the Keplerian disk terminates at the 
last stable orbit and yet the magnetosphere is pushed farther inward. 
The threshold is therefore dependent of the magnetic field strength 
--- the stronger the magnetic field the higher the threshold. This is 
certainly in agreement with the atoll/Z paradigm, but we argue that it 
is also generally true, even for individual sources within each (atoll 
or Z) category. For atoll sources, the kHz QPOs also seem to vanish at 
low accretion rate. Perhaps the ``disengagement'' between the 
magnetosphere and the Keplerian disk also takes place under such 
circumstances, because of, for instance, the presence of quasi-spherical 
advection-dominated accretion flow (ADAF) close to the neutron star. 
Unfortunately, in this case, the estimation of the accretion rate 
threshold would require a knowledge of the physical mechanisms that 
cause the disengagement. If the ADAF is responsible, the threshold is 
likely dependent of the magnetic field of the neutron star.
\end{abstract}

\keywords{accretion, accretion disks --- binaries: general --- 
relativity --- stars: oscillations --- X-ray: stars}

\section{Introduction}
The detection of kilohertz quasi-periodic oscillation (kHz QPO) in
low-mass X-ray binaries is arguably the greatest discovery that the 
{\it Rossi X-ray Timing Explorer} (RXTE) has made to date. Such
signals almost certainly originate in the immediate vicinity of
central neutron stars, given that the fastest oscillations are 
observed to occur on dynamical timescales near such objects (review 
by van der Klis 2000). The prospect of using kHz QPOs to probe the 
effects of strong gravity near neutron stars is therefore very 
exciting. 

Various models (e.g., Klein et al. 1996; Miller et at. 1998; Stella \& 
Vietri 1999; Osherovich \& Titarchuk 1999) have been proposed to 
explain kHz QPOs that, in nearly all cases, come in pairs (note that 
we choose not to discuss the QPOs observed during thermonuclear
bursts). Except for the ``photon bubble model'' (Klein et al. 1996),
all other models invariably associate one of the pair to the Keplerian 
motion of clumps of matter or ``hot spots'' at the inner edge of a 
geometrically thin accretion disk (although the ``sonic point model'' 
differs in detail, e.g., Miller et at. 1998; however, see Lai
1998). The inner edge of the disk is determined by the pressure
balance between the accreted matter and the magnetic field of the 
neutron star. As the mass accretion rate increases, the ram pressure
of the accreted matter increases, which squeezes the magnetosphere
more and thus the accretion disk extends farther toward the neutron
star. Therefore, we expect that the frequency of the ``Keplerian QPO'' 
increases with the accretion rate, which agrees with the observations 
(van der Klis 2000). Since no Keplerian flow can exist inside the last 
stable orbit, we expect that there is an upper limit to the frequency
of this QPO --- any further increase in the accretion rate cannot
result in an increase in the frequency of the QPO. Observing this
limit would provide strong observational evidence for the presence of 
the last stable orbit around neutron stars, a natural consequence of 
strong gravity. 

However, due to the lack of detailed knowledge on how the signals are 
produced in the first place, it is still not clear how kHz QPOs would 
behave when the accretion disk reaches the last stable orbit. Without 
considering any physical mechanisms responsible for modulating 
the X-ray emission, one might take it for granted that the frequency 
of the Keplerian QPO would saturate at sufficiently high accretion 
rate (e.g., Zhang et al. 1998b; Lai 1998). On the other hand, Cui et 
al. (1998) emphasized the importance of disk-magnetosphere 
interaction in producing the QPOs, based on
a detailed study of the evolution of a kHz QPO observed of Aquila X-1, 
a transient atoll source, throughout the rising phase of an X-ray 
outburst. In this Letter, we generalize the ideas proposed by Cui 
et al. to all kHz QPO sources. 

\section{Saturation of kHz QPO Frequency at High X-ray Flux?}
Using data from the RXTE observations of 4U 1820-30, which spanned
over almost one year, Zhang et al. (1998b) found that the kHz QPOs 
detected in the source evolved strongly in frequency at low X-ray 
fluxes but reached a plateau at high fluxes. They attributed such a 
saturation in the QPO frequency to the presence of the last stable 
orbit. In other words, the plateau frequency of one of the kHz QPOs 
would represent the Keplerian frequency at the last stable orbit, if 
the interpretation is correct. 

However, for atoll or Z sources, the X-ray flux is not always a 
reliable indicator for the mass accretion rate (e.g., M\'{e}ndez et
al. 1999; M\'{e}ndez 1999; Ford et al. 2000), which fundamentally 
determines the magnetospheric radius of the neutron star and thus the 
location of the inner edge of the accretion disk. Moreover, it is now 
well established that the frequency of kHz QPOs is {\em not} well 
correlated with X-ray flux over a long period of time, although 
short-term linear correlations seem to be present (Zhang et al. 1998a; 
M\'{e}ndez 1999 and references therein). Therefore, for 4U 1820-30 
the observed evolution of the kHz QPOs with X-ray flux may not be a 
good representation of that with accretion rate. The saturation of 
the QPO frequencies could simply be an artifact caused by poor 
sampling of the data (M\'{e}ndez 1999, private communication). 
Unfortunately, there is no reliable way of determining (or even 
quantifying) the accretion 
rate for atoll or Z sources at present. One empirical approach that is
often taken is to use the position long the distinctive tracks in the 
color-color or color-intensity diagrams (which define atoll or Z 
sources) as a qualitative measure of the accretion rate (van der
Klis 1995). The approach seems effective since the frequency of kHz 
QPOs is shown to be uniquely correlated with such ``color track
position'' for a number of sources (M\'{e}ndez et al. 1999; 
M\'{e}ndez 1999; van Straaten et al. 2000). 

Kaaret et al. (1999) re-analyzed the data of Zhang et al. (1998b) and
added data from more recent observations. They seemed to confirm that 
the QPO frequencies leveled off at high X-ray fluxes, although this 
result was {\em not} reproduced in other studies (van der Klis 2000). 
When they plotted the frequencies against the inferred accretion rate 
(as indicated by the color track position in the color-intensity 
diagram that they defined), however, the frequencies varied {\em
monotonically} over the entire range. The authors noted an apparent 
deviation in the correlation at high accretion rates, with very 
limited dynamical range, from that extrapolated from low accretion 
rates. They then interpreted the deviation as evidence for the last 
stable orbit. We would like to note, however, that the color track 
position of the source is only an {\em empirical} indicator for the 
accretion rate. It is not clear at all how the two are 
{\em quantitatively} related; there is certainly no compelling reason 
to expect that they are necessarily {\em linearly} correlated. 
Consequently, the correlation between the QPO frequency and the color 
track position is expected to be complex (but is not known). A more 
serious question regarding the interpretation is why no other sources 
show a similar behavior (i.e., the claimed saturation of kHz QPO 
frequencies at high accretion rate), if the phenomenon were to 
originate in something so fundamental. It seems unlikely that the 
uniqueness of 4U~1820-30 can be the answer to this question.

\section{Clues from Transient Atoll Sources}
Transient atoll sources are nearly ideal systems for a detailed study
of the evolution of kHz QPOs with mass accretion rate, because of the
large dynamical range that they provide during an X-ray outburst. 
Unfortunately, they are very few in number, and even fewer that undergo
outbursts on time scales short enough that they can be be observed. 
Nevertheless, high-quality data is available for some of these sources.

Aql X-1 is known for experiencing frequent outbursts (van Paradijs \&
McClintock 1995). During an outburst in 1998, the source was
intensively monitored throughout the rising phase of the outburst (Cui
et al. 1998). Although the origin of X-ray outbursts is generally
unknown, there appears to be some consensus now that thermal 
instability causes a sudden surge in the mass accretion rate through 
the disk and thus initiates an X-ray outburst (review by King 1995). 
Therefore, we know, at least qualitatively, how the mass accretion
rate evolves during an outburst. Cui et al. found that the rising 
phase seems to be quite simple (at least for Aql X-1): the X-ray flux 
correlates fairly well with the color track position, compared to the 
decaying phase where no correlation exists between the two quantities 
on long time scales (Zhang et al. 1998a; M\'{e}ndez 1999). Perhaps, 
during the rising phase of an outburst, the X-ray flux is simply 
proportional to the mass accretion rate, as usually expected. 

The evolution of a kHz QPO (not a pair) detected in Aql X-1 was 
carefully followed during the rising phase of the outburst (Cui et
al. 1998). The QPO was not detected at the beginning of the outburst, 
when the accretion rate was presumably low; it then appeared and 
persisted through the intermediate range of the accretion rate; and 
it vanished again when the accretion rate exceeded a certain threshold 
($\dot{M}_h$) near the peak of the outburst. To account for such an 
evolution of the kHz QPO, Cui et al. proposed the following physical 
scenario, as illustrated in Figure~1. 

In the quiescent state, the mass accretion takes place in the form 
of advection-dominated accretion flow (ADAF) close to the central 
neutron star, and in the forms of a standard Keplerian disk farther 
away (e.g., Narayan \& Yi 1994, 1995; see also Menou et al. 1999 for 
discussions of neutron star systems). Therefore, there is no direct 
interaction between the Keplerian disk and the magnetosphere of the 
neutron star in this
state. Such an interaction was argued to be essential for producing 
X-ray modulation associated with kHz QPOs, consequently, there would 
be no kHz QPOs in the quiescent state. At the onset of an X-ray outburst, 
the mass accretion rate begins to increase rapidly, so the ADAF region
shrinks and the Keplerian disk moves inward (Narayan 1997). As 
soon as the disk starts to interact with the magnetosphere, kHz QPOs 
are produced. The accretion rate continues to increase as the 
outburst proceeds, so the magnetosphere is pushed farther toward the 
neutron star and the disk extends farther inward. After the disk has 
reached the last stable orbit, any increase in the mass accretion rate 
disengages the disk from the magnetosphere, causing the QPOs to
disappear. For Aql X-1, assuming the X-ray flux is proportional to the 
accretion rate, the inferred low and upper limits on the magnetic
field strength are certainly consistent with our expectation of an
atoll source (Cui et al. 1998). 

Another transient atoll source, 4U 1608-52, showed a remarkably
similar pattern, based on the inferred mass accretion rate from the 
color-color diagram, in the evolution of its kHz QPO during the 
decaying phase of an outburst (M\'{e}ndez et al. 1999). Therefore, we 
propose that {\em it is the disappearance of kHz QPOs at high mass 
accretion rate that provides evidence for the presence of the last 
stable orbit.} 

\section{Application to All kHz Sources}
Without an exception, the kHz QPOs vanish at high mass accretion rate
for all sources (van der Klis 2000). For different sources, however,
this seems to occur at a different accretion rate. For instance, 
$\dot{M}_h$ is certainly higher for Z sources than that
for atoll sources, perhaps suggesting a critical role that the
magnetic field plays (according to the atoll/Z paradigm; van der Klis
1995). This would be naturally explained by our model since the model
requires that the stronger the magnetic field the higher the threshold 
(see \S~5 for a quantitative treatment). Moreover, extending the model 
to all kHz sources would solve a long-standing observational puzzle as 
to why the kHz QPOs disappear at a higher accretion rate for a brighter 
source (van der Klis 2000). It is interesting to note that no kHz QPOs 
have ever been detected in a group of bright atoll sources (GX 3+1, 
GX 9+1, GX 13+1, and GX 9+9; Wijnands et al. 1998; Strohmayer 1998; 
Homan et al. 1998). Perhaps, for these sources, the accretion rate 
always remains above $\dot{M}_h$. 

At low mass accretion rates, with the exception of 4U 1728-34, kHz
QPOs become undetectable when the accretion rate is below a 
certain threshold ($\dot{M}_l$) for all atoll sources; for Z sources, 
on the other hand, the QPOs are detected down to the lowest inferred 
mass accretion rate (van der Klis 2000). Perhaps, the ADAF scenario 
can also be applied to persistent atoll sources or even Z sources. 
If so, $\dot{M}_l$ would likely depend sensitively on the magnetic 
field strength, as we will demonstrate in the next section. 

\section{Disk--Magnetosphere Interaction}
One of the critical issues that the kHz QPO models do not explicitly 
address is the physical mechanism that modulates the X-ray emission. 
The models are presently still at the level of simply associating the 
natural frequencies in a low-mass neutron star binary 
to the kHz QPOs observed. However, a mechanism is clearly needed for 
the natural frequencies to manifest themselves observationally. An 
obvious candidate is the interaction between the Keplerian disk and 
the magnetosphere of the neutron star, as was proposed in the 
original formulation of the beat-frequency model (Alpar \& Shaham 
1985). Such an interaction can create inhomogeneity or warping in the 
accretion flow (Vietri \& Stella 1998; Lai 1999), which may cause 
X-ray emission to be modulated at the orbital frequency.

Assuming a dipole field, the radius of the magnetosphere, in units of
the Schwarzschild radius ($\equiv 2GM/c^2$), is approximately given by 
(Frank et al. 1992):
\begin{equation}
r_m \approx 2.19 \dot{m}^{-2/7} m^{-10/7} B^{4/7}_8 r^{12/7}_6,
\end{equation}
where m is the mass of the neutron star in solar units, $\dot{m}$ is
the mass accretion rate in Eddington units ($1.39 \times 10^{18} m 
\mbox{ }g\mbox{ }s^{-1}$), $B_8$ is the dipole field of the neutron
star in units of $10^8$ G, and $r_6$ is the radius of the neutron star 
in units of $10^6$ cm. To derive $\dot{M}_h$, therefore, we set 
$r_m = 3$ (ignoring the effects of slow rotation of the neutron star). 
We have
\begin{equation}
\dot{M}_h = 0.33 m^{-5} B^2_8 r^6_6.
\end{equation}
For neutron stars with $m = 2$ and $r_6 = 1$, $\dot{M}_h \approx 0.01$ 
for $B = 10^8$ G or $\sim 1$ for $B = 10^9$ G, which are roughly what 
we expect of atoll or Z sources, respectively (van der Klis 1995).

At low mass accretion rates, if the ADAF is responsible for truncating 
the Keplerian disk, we can derive $\dot{M}_l$ by setting $r_m = r_{tr}$, 
where $r_{tr}$ is the radius at which the Keplerian disk makes a 
transition to the quasi-spherical ADAF. However, the physical origin 
of the transition is still poorly understood. Observationally,
$r_{tr}$ is clearly a function of $\dot{m}$ --- the higher the mass
accretion rate the smaller the transition radius (Narayan 
1997). Assuming $r_{tr} \propto \dot{m}^{-\alpha}$ (where 
$\alpha > 0$), equating $r_{tr}$ to $r_m$ yields
\begin{equation}
\dot{M}^{\alpha - 2/7}_l \propto m^{10/7} B^{-4/7}_8 r^{-12/7}_6.
\end{equation}
If $\alpha > 2/7$, the stronger the magnetic field the lower
$\dot{M}_l$ will be; the converse is true for $\alpha < 2/7$. The 
former appears to be
supported by observations, given that the kHz QPOs persist down the
the lowest mass accretion rate for Z sources while they seem to 
disappear at some point for atoll sources. Of course, the ADAF 
scenario might break
down entirely for Z sources; i.e., the mass accretion process 
always takes the form of a Keplerian disk, so the kHz QPOs are always
detectable at low mass accretion rates. 

Meyer \& Meyer-Hofmeister (1994) suggested that a cool Keplerian disk 
could undergo a phase transition to a hot, quasi-spherical corona, when 
the gas density in the transition region becomes too low to effectively 
radiate away the energy released during the mass accretion process. 
If such a phase transition is relevant for ADAF, then 
$r_{tr} = 18.3 m^{0.17/1.17} \dot{m}^{-1/1.17}$ (i.e., $\alpha = 0.84
> 2/7$; Liu et al. 1999). We would have
\begin{equation}
\dot{M}_l = 41.7 m^{2.76} B^{-1}_8 r^{-3}_6,
\end{equation}
which would likely be super-Eddington for atoll or Z
sources. Therefore, it is not clear if the model can be applied to 
neutron star systems.

\section{Summary}
The observations of the kHz QPO phenomenon seem to suggest that the 
interaction between the Keplerian accretion disk and the magnetosphere 
of the neutron star is directly responsible for modulating the X-ray 
emission. For a given source, the presence (or absence) of such an 
interaction dictates the appearance (or disappearance) of the QPOs. At
high mass accretion rate, we argue that the presence of the last
stable orbit manifests itself in the disappearance of the QPOs, as
opposed to the saturation in the QPO frequency. The difference may
only seem semantic since, quantitatively, both interpretations require 
that the neutron star is inside the last stable orbit. However, we 
feel that it is imperative for the models to begin to address such 
critical issues as modulation mechanisms for kHz QPOs, in light of the 
ever improving quality of the data. 

One critical question is whether the neutron star magnetosphere can be 
disengaged from the Keplerian disk at the last stable orbit. Studies 
have shown that the evolution of the magnetic field configuration is 
very complicated as the disk approaches the last stable orbit (e.g., 
Lai 1998), but the exact solution is not known at
present. Intuitively, as the accretion process proceeds from
the inner edge of the Keplerian disk onto the surface of the neutron 
star, the ram pressure of the accreted matter continues to push the 
magnetosphere inward. In this case, the
magnetosphere only interacts with the non-Keplerian accretion flow in
the ``gap'' between the last stable orbit and the neutron star surface,
but {\em not} directly with the Keplerian flow in the disk. The
importance of such gap accretion has been studied extensively
(Klu\`{z}niak \& Wagoner 1985; Klu\`{z}niak \& Wilson 1991).

The situation is less certain at low mass accretion rates. In fact,
observationally it can still be argued whether the QPOs actually 
disappear, given that in nearly all cases the upper limits derived 
are comparable to the fractional rms amplitudes of the QPOs measured 
at high accretion rates (M\'{e}ndez 2000, private communication). In 
the case of 4U 0614+09, however, the 95\% upper limit is only about half 
of the measured amplitude when the source is bright (M\'{e}ndez et al. 
1997). Therefore, we have at least one source in which the QPOs, if 
present at all, are definitely much weaker at low accretion rates 
(i.e., below $\dot{M}_l$). Moreover, we note that often the kHz QPOs 
{\em strengthen} (relative to the average source intensity) as the 
accretion rate {\em decreases} (Wijnands et al. 1997; Wijnands \& 
van der Klis 1997; Smale, Zhang, \& White 1997). If the QPOs do 
disappear at low accretion rate, some physical process, like the 
ADAF, could be present to truncate the Keplerian disk at large distance 
from the neutron star under such circumstances. This would destroy 
the disk-magnetosphere interaction and thus the kHz QPOs for atoll
sources. The process may also operate in Z sources: the persistence 
of the kHz QPOs in such cases can be attributed to a lower accretion 
rate threshold that is due to a stronger magnetic field of the neutron 
star. Alternatively, the process bears no relevance to Z sources. The 
disk-magnetosphere interaction is always present at low accretion 
rates, so are the QPOs. 

\acknowledgments
I thank Dong Lai for a stimulating discussion on the subject. Part of
this work was completed when I was attending the workshop on ``X-ray 
Probes of Relativistic Astrophysics'' at the Aspen Center for Physics 
(ACP) in the summer of 1999. I acknowledge helpful discussions with 
the participants of the workshop, in particular, with Michiel van der 
Klis and Mariano M\'{e}ndez on several important observational
issues. I also wish to acknowledge financial assistance from the ACP. 
This work was supported in part by NASA grants NAG5-7990 and NAG5-7484.

\clearpage
\begin{figure}
\psfig{figure=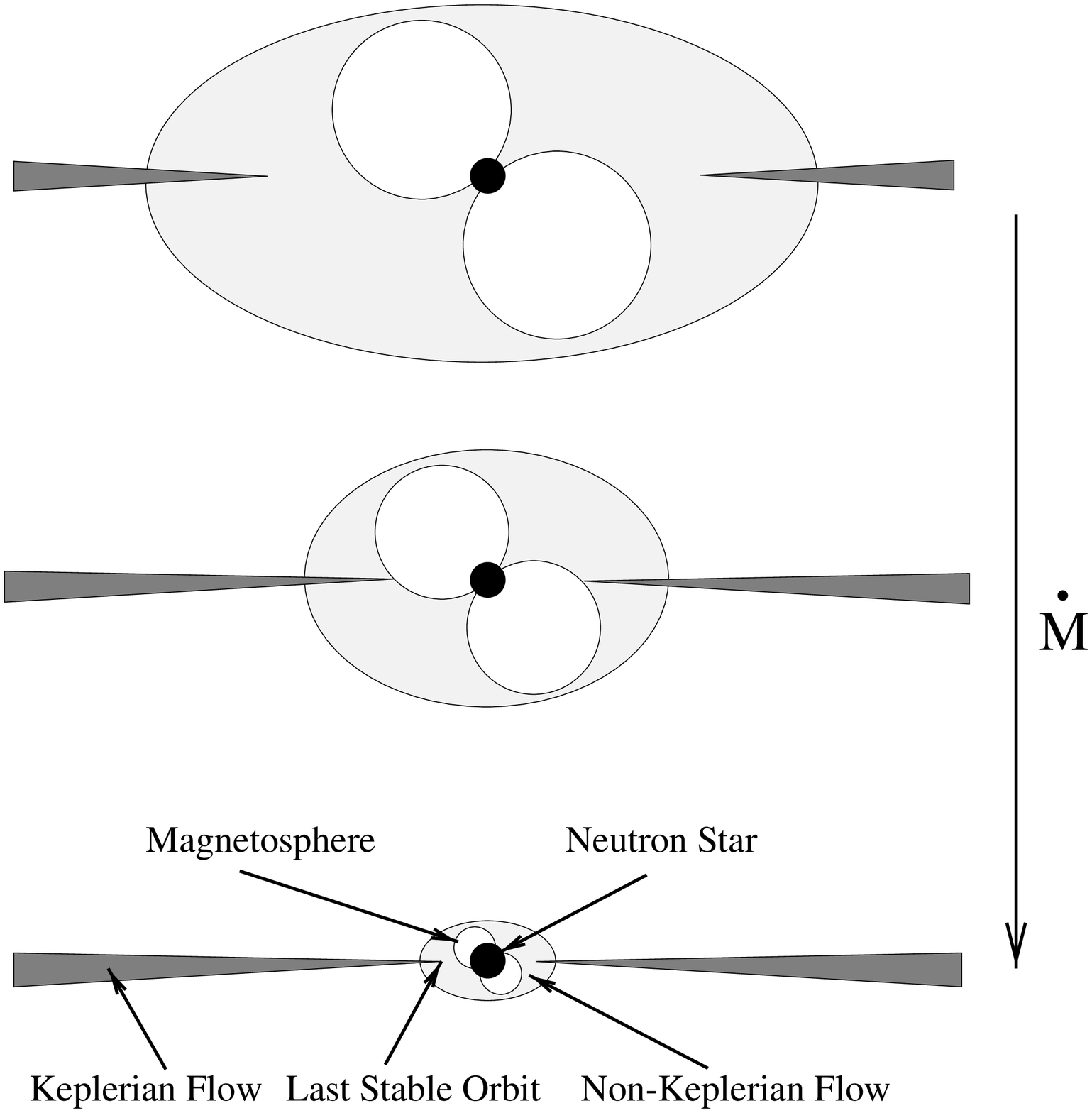,width=6in}
\caption{Schematic illustration of the geometry and evolution of
accretion flows in a low-mass X-ray binary (not to scale). In the
proposed scenario (see text), the inner edge of the Keplerian disk
pushes the magnetosphere of the neutron star inward and the region 
of non-Keplerian flow (e.g., ADAF) becomes smaller, as the mass 
accretion rate increases (from top to bottom). Note that the kHz 
QPOs are produced when the disk begins to interact with the 
magnetosphere, and they persist until the disk reaches the last 
stable orbit. }
\end{figure}

\clearpage

\begin{references}
\reference{} Alpar,~M.~A., \& Shaham,~J. 1985, {\em Nature}, 316, 239
\reference{} Cui,~W., Barret,~D., Zhang,~S.~N., Chen~W., Boirin,~L.,
\& Swank,~J. 1998, \apjl, 502, L49
\reference{} Ford,~E.~C., et al. 2000, \apj, submitted
(astro-ph/0002074)
\reference{} Frank,~J., King,~A.~R., \& Raine,~D. 1992, Accretion
Power in Astrophysics (Cambridge: Cambridge University Press)
\reference{} Homan,~J., van der Klis,~M., Wijnands,~R., Vaughan,~B., 
\& Kuulkers,~E. 1998, \apj, 499, L41
\reference{} Kaaret,~P., Piraino,~S., Bloser,~P.~F., Ford,~E.~C., \&
Grindlay,~J.~E. 1999, \apjl, 520, L37
\reference{} Klein,~R.~I., Jernigan,~J.~G., Arons,~J., Morgan,~E.~H.,
\& Zhang,W. 1996, \apj, 469, L85
\reference{} King,~A.~R. 1995, in ``X-ray Binaries'', eds. W. H. G. Lewin, J. van Paradijs, \& E. P. J. van den Heuvel (Cambridge U. Press, Cambridge) p. 419
\reference{} Klu\`{z}niak,~W., \& Wagoner,~R.,~V. 1985, \apj, 297, 548 
\reference{} Klu\`{z}niak,~W., \& Wilson,~J.~R. 1991, \apjl, 372, L87
\reference{} Lai,~D. 1998, \apj, 502, 721
\reference{} Lai,~D. 1999, \apj, 524, 1030
\reference{} Liu,~B.~F., Yuan,~W., Meyer,~F., Meyer-Hofmeister,~E., \&
Xie,~G.~Z. 1999, \apjl, 527, L17
\reference{} M\'{e}ndez,~M., van der Klis,~M., Ford,~E.~C.,
Wijnands,~R., van Paradijs,~J. 1999, \apjl, 511, L49
\reference{} M\'endez, M. 2000, Proc. 19th Texas Symposium on Relativistic
Astrophysics and Cosmology, ed. J. Paul, T. Montmerle, \&
E. Aubourg (Amsterdam: Elsevier), in press (astro-ph/9903469)
\reference{} M\'{e}ndez,~M., et al. 1997, \apj, 485, L37
\reference{} Menou,~K., Esin,~A.~A., Narayan,~R., Garcia,~M.~R.,
Lasota,~J.-P., \& McClintock,~J.~E. 1999, \apj, 520, 276
\reference{} Meyer,~F., \& Meyer-Hofmeister,~E. 1994, \aap, 288 175
\reference{} Miller,~M.~C., Lamb,~F.~K., \& Psaltis,~D., \apj, 508, 791
\reference{} Narayan,~R., \& Yi,~I. 1994, \apjl, 428, L13
\reference{} Narayan,~R., \& Yi,~I. 1995, \apj, 444, 231
\reference{} Narayan,~R. 1997, in ``Accretion Phenomena and Related
Outflows'', eds D.~T.~Wickramasinghe, G.~V.~Bicknell, \& L.~Ferrario
(ASP Conf. Series Vol. 121; San Francisco: ASP), p. 75
\reference{} Osherovish,~V., \& Titarchuk,~L., 1999, \apj, 522, L113
\reference{} Smale,~A.~P., Zhang,~W., \& White,~N.~E. 1997, \iaucirc\ 6507
\reference{} Stella,~L., \& Vietri,~M. 1999, Phys. Rev. Lett. 82, 17
\reference{} Strohmayer,~T. 1998, in "Some Like It Hot", AIP
Conf. Proc. 431, 397 
\reference{} van der Klis,~M. 1995, in ``X-ray Binaries'', eds. W. H. G. Lewin, J. van Paradijs, \& E. P. J. van den Heuvel (Cambridge U. Press, Cambridge) p. 252
\reference{} van der Klis,~M. 2000, \araa, submitted
(astro-ph/0001167)
\reference{} van Paradijs,~J., \& McClintock,~J.~E. 1995, in ``X-ray Binaries'', eds. W. H. G. Lewin, J. van Paradijs, \& E. P. J. van den Heuvel (Cambridge U. Press, Cambridge) p. 58
\reference{} van Straaten,~S., Ford,~E.~C., van der Klis,~M.,
M\'{e}ndez,~M., \& Kaaret,~P. 2000, \apj, in press (astro-ph/0001480)
\reference{} Vietri,~M., \& Stella,~L. 1998, \apj, 503, 350
\reference{} Wijnands,~R., van der Klis,~M., \& van Paradijs,~J. 1998,
in ``The Hot Universe'', IAU Symp. 188, p. 370
\reference{} Wijnands,~R., van der Klis,~M., van Paradijs,~J., 
Lewin,~W.~H.~G., Lamb,~F.~K., Vaughan,~B.~A., \& Kuulkers,~E. 1997,
\apj, 479, L141 
\reference{} Wijnands,~R., \& van der Klis,~M. 1997, \apj, 482, L65 
\reference{} Zhang,~W., Jahoda,~K., Kelley,~R.~L., Strohmayer,~T.~E., 
Swank,~J.~H., \& Zhang,~S.~N. 1998a, \apj, 495, L9
\reference{} Zhang,~W., Smale,~A.~P., Strohmayer,~T.~E., \&
Swank,~J.~H. 1998b, \apjl, L171
\end{references}
\end{document}